\documentclass[letterpaper, 10 pt, conference]{ieeeconf}  
\IEEEoverridecommandlockouts                         
\usepackage[T1]{fontenc}
\usepackage{aecompl}
\usepackage{epstopdf}
\usepackage{cite}
\usepackage{amsmath,amssymb,amsfonts}
\usepackage{algorithmic}
\usepackage{algorithm}
\usepackage{graphicx}
\usepackage{textcomp}
\usepackage{xcolor}
\usepackage{float}
\usepackage{graphicx}
\usepackage{picinpar}
\usepackage{babel}
\usepackage{amsmath}
\usepackage{url}
\usepackage[latin1]{inputenc}
\usepackage{colortbl}
\usepackage{soul}
\usepackage{multirow}
\usepackage{pifont}
\usepackage{color}
\usepackage{alltt}
\usepackage[hidelinks]{hyperref}
\usepackage{enumerate}
\usepackage{siunitx}
\usepackage{breakurl}
\usepackage{epstopdf}
\usepackage{pbox}
\usepackage{authblk}
\usepackage{babel}
\usepackage{mathrsfs,balance,subfigure}
\usepackage{tcolorbox}
\usepackage{fancyhdr}

\title{\LARGE \bf
Performance-Driven Controller Tuning via Derivative-Free Reinforcement Learning
}

\author{Yuheng Lei$^{1}$, Jianyu Chen$^{2\ 3}$, Shengbo Eben Li$^{1*}$, Sifa Zheng$^{1}$
\thanks{This study is supported by National Key R\&D Program of China under 2020YFB1600200 and sponsored by Tsinghua-Toyota Joint Research Fund.}
\thanks{$^{1}$School of Vehicle and Mobility, Tsinghua University, Beijing, China}%
\thanks{$^{2}$Institute for Interdisciplinary Information Sciences, Tsinghua University, Beijing, China}
\thanks{$^{3}$Shanghai Qi Zhi Institute, Shanghai, China}%
\thanks{$^{*}$All correspondence should be sent to Shengbo Eben Li with email: {\tt\small lishbo@tsinghua.edu.cn}}
}

\begin{document}

\maketitle
\thispagestyle{empty}
\pagestyle{empty}

\thispagestyle{fancy}
\fancyhead{}
\lhead{}
\lfoot{\footnotesize{\textcopyright 2022 IEEE. Personal use of this material is permitted. Permission from IEEE must be obtained for all other uses, in any current or future media, including reprinting/republishing this material for advertising or promotional purposes, creating new collective works, for resale or redistribution to servers or lists, or reuse of any copyrighted component of this work in other works.}}
\cfoot{}
\rfoot{}

\begin{abstract}

Choosing an appropriate parameter set for the designed controller is critical for the final performance but usually requires a tedious and careful tuning process, which implies a strong need for automatic tuning methods. 
However, among existing methods, derivative-free ones suffer from poor scalability or low efficiency, while gradient-based ones are often unavailable due to possibly non-differentiable controller structure.
To resolve the issues, we tackle the controller tuning problem using a novel derivative-free reinforcement learning (RL) framework, which performs timestep-wise perturbation in parameter space during experience collection and integrates derivative-free policy updates into the advanced actor-critic RL architecture to achieve high versatility and efficiency.
To demonstrate the framework's efficacy, we conduct numerical experiments on two concrete examples from autonomous driving, namely, adaptive cruise control with PID controller and trajectory tracking with MPC controller. 
Experimental results show that the proposed method outperforms popular baselines and highlight its strong potential for controller tuning.

\end{abstract}
\section{INTRODUCTION}
\label{sec:intro}

Controller tuning is an indispensable process in industries that directly decides the overall control performance. 
Manual tuning through trial-and-error and exhaustive grid search are the most common and straightforward strategies.
However, they can be highly time-consuming and labor-intensive, and usually result in sub-optimal controllers due to limited experiments, especially when a large number of parameters and large-scale complex dynamic systems are involved \cite{neumann2019data, wang2021learning, marco2016automatic}. 
Although some empirical guidelines have been summarized for different controller types (e.g., \cite{o2009handbook} for PID, \cite{garriga2010model} for MPC), considerable expertise and tedious efforts are still required to find appropriate parameter settings. 
Therefore, automatic controller tuning methods are greatly desired to solve the difficulties and have drawn substantial attention from researchers.

In many optimal control schemes (e.g., LQR and MPC), a simplified dynamics model of the real system is required. 
A typical way is to first conduct model identification via open-loop experiments and then design the corresponding model-based controller \cite{xu2019automated}. 
However, it separates the two stages apart and may suffer from the objective mismatch between training the dynamics model to minimize prediction errors and tuning the controller towards improved performance on the targeted control task \cite{lambert2020objective} and may lead to sub-optimal control performance \cite{bansal2017goal}. 
The problem is particularly amplified when the model capacity is limited with respect to the high dimensional observation space (e.g., images), and the awareness of downstream control tasks is critical to learning \cite{nair2020goal}. 
This gives rise to an alternative paradigm called \textit{Identification for Control} (I4C) that jointly optimizes the model and controller according to the desired closed-loop performance on the given task \cite{gevers2005identification}. 
Following the I4C paradigm, recently many researches have shown the possibility to train models without explicitly minimizing the model error \cite{piga2019performance, bansal2017goal, freeman2019learning, risi2019deep}. 

\begin{figure}[t]
\centering
\subfigure[Derivative-free optimization]{\includegraphics[width=0.45\linewidth, keepaspectratio=true,trim=0 170 0 0,clip]{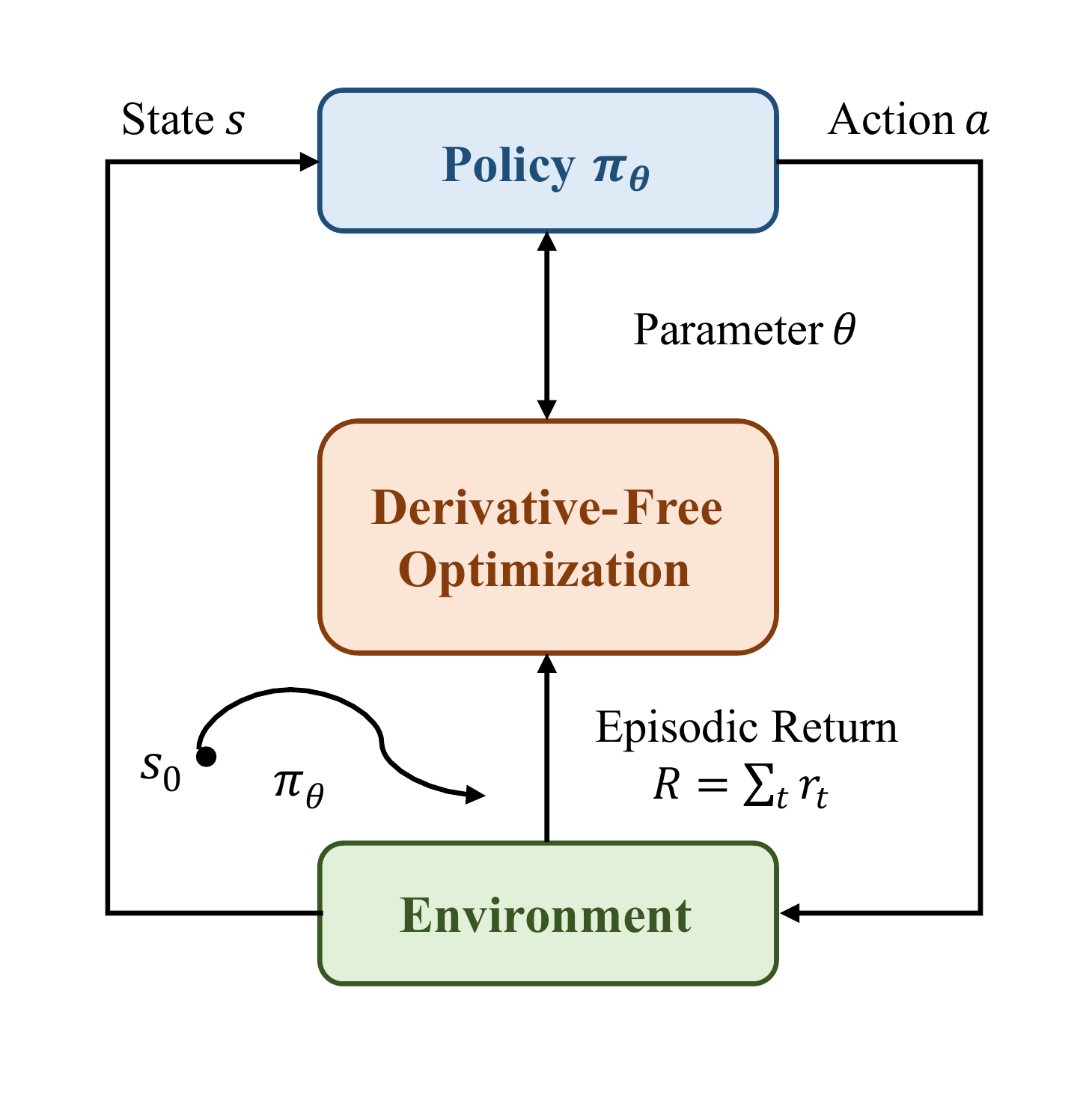}}
\subfigure[Reinforcement learning]{\includegraphics[width=0.47\linewidth, keepaspectratio=true,trim=0 150 0 0,clip]{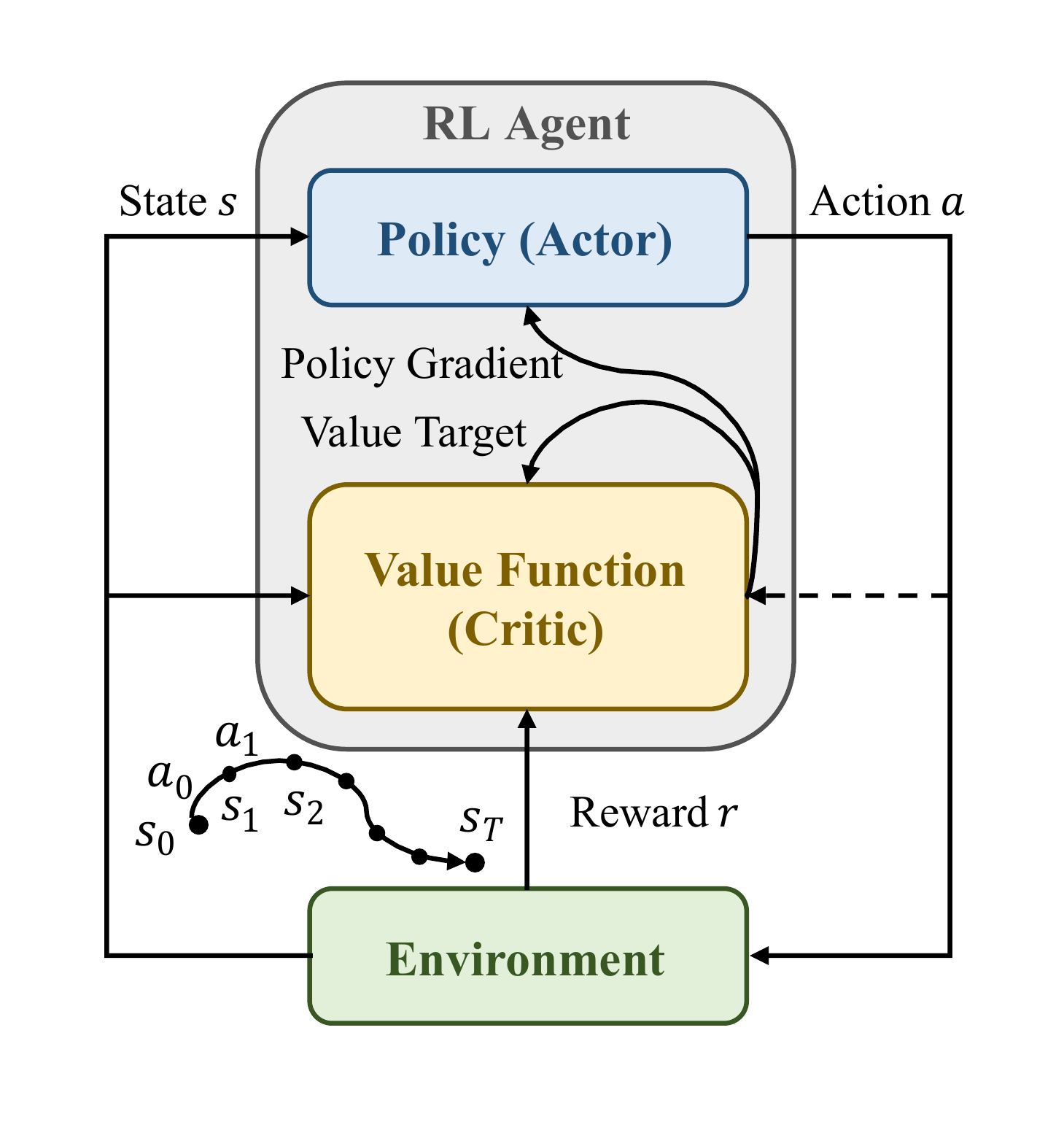}}
\caption{Two alternative approaches to controller tuning. A crucial difference in reinforcement learning is the formulation of sequential decision problem and temporal credit assignment.}
\vspace{-0.1in}
\label{fig:alg}
\end{figure}

Since there is no apparent connection between the parameters and the underlying performance objective, a natural choice is to use \textit{derivative-free optimization} (DFO, a.k.a., \textit{zeroth-order optimization} or \textit{black-box optimization}) methods. 
\textit{Bayesian optimization} (BO) is widely used to tune controllers in the community, e.g., closed-loop poles \cite{neumann2019data}, PID gains \cite{wang2021learning}, dynamics model \cite{bansal2017goal, piga2019performance}, cost function and horizon length in LQR/MPC \cite{marco2016automatic, lucchini2020torque}.
BO usually performs well on low-dimensional problems (typically smaller than 20) if the probabilistic prior and the acquisition function are carefully designed, albeit usually not a trivial task \cite{shahriari2015taking}. 
BO is notoriously difficult to scale to high dimensions; for instance, \cite{roveda2020two} splits the tuning of equivalent mass parameters and PID gains into two stages to reduce the complexity. 
The family of \textit{evolutionary algorithms} (EAs) is another popular choice, including GA \cite{risi2019deep}, ES \cite{salimans2017evolution}, CMA-ES \cite{jain2021optimal}, ARS \cite{mania2018simple, jain2021pixels, huang2021accelerated}.
Contrary to BO, evolutionary algorithms do not rely on probabilistic priors and have shown promising results on complex high-dimensional problems, e.g., training neural networks for vision-based control \cite{risi2019deep, jain2021pixels}. 
Together with some other advantages, including diverse exploration and high robustness, EAs have regained popularity in recent years. 
However, in such methods, a large number of samples are needed to perform an update, which often results in drastically high sample complexity.

Alternatively, as most of the optimal control problems can be represented as a \textit{Markov Decision Process} (MDP), \textit{reinforcement learning} (RL) is an effective tool to leverage \cite{sutton2018reinforcement}.
From the perspective of RL, the aforementioned I4C paradigm can also be interpreted as training the control policy end-to-end with model-free methods.
However, existing policy gradient-based RL algorithms need to differentiate through the control policy, which is applicable for neural networks but usually not trivial for classic controllers.
Few exceptional examples include linear feedback controllers \cite{matni2019self}, specific sub-classes of real-time optimization-based controllers \cite{amos2018differentiable, agrawal2020learning}, etc. 
Take the latter as an example, if some certain assumptions hold, the derivatives of the output solution with respect to the design parameters can be computed analytically by differentiating the KKT optimality conditions \cite{buskens2001sensitivity}. 
Some recent research have tried to integrate the analytical derivatives of controllers into the common RL framework \cite{gros2019data, gros2021reinforcement}, which usually leads to higher sample efficiency by utilizing the sequential nature of the problem.
However, since the prerequisites are not always satisfied in practical use, the derivatives may be misleading (e.g., when not reaching a fixed point of the optimization problem \cite{amos2018differentiable}) or intractable (e.g., non-differentiable cost and dynamics). 
Thus, such approaches are still highly restricted and do not generalize well as derivative-free optimization.
Besides, existing RL methods usually explore by adding stochastic noise on the output action independently per timestep. 
This mechanism could be problematic on real systems since the resulting jittery trajectories lead to increased wear-and-tear, and also poor exploration given that most systems own low-pass-filtering characteristics\cite{raffin2022smooth}.

In this paper, we propose a novel derivative-free reinforcement learning framework called zeroth-order actor-critic (ZOAC) for performance-driven controller tuning problem, which makes several contributions more beneficial than prior arts.
Firstly, the framework formulates the controller tuning problem as an MDP to leverage the sequential nature for higher efficiency.
Secondly, it unifies first-order policy evaluation (PEV) and zeroth-order policy improvement (PIM) into an advanced on-policy actor-critic architecture, which ensures its versatility for various types of controllers. The timestep-wise perturbation in parameter space also leads to better exploration.
To demonstrate the framework's efficacy, we present the results of numerical experiments on two controller tuning examples: tuning a PID controller for adaptive cruise control and tuning an MPC controller for trajectory tracking.  
Experimental results show that the parameters of controllers can be optimized via the proposed auto-tuning framework more efficiently than existing popular baselines, highlighting its strong potential for automatic controller tuning in industries.


\section{PRELIMINARIES}
\label{sec:prelim}

\subsection{Evolutionary Strategies}
\label{sec:prelim.zog}
Let $f(\theta)$ denotes a function to be optimized, which is possibly non-smooth, non-differentiable and only zeroth-order information (i.e., the function output itself) is available. Evolution strategies (ES) \cite{salimans2017evolution, mania2018simple, jain2021pixels}, one of the popular evolutionary algorithms,  optimizes a Gaussian smoothed objective $\mathbb{E}_{\epsilon\sim\mathcal{N}(0,I)}[f(\theta+\sigma\epsilon)]$,
where $\sigma$ is the standard deviation. The zeroth-order gradient can be derived using the log-likelihood ratio trick and the probability density function of Gaussian distribution:
\begin{equation}
\label{eq:esgrad}
\nabla_\theta f(\theta)=\frac{1}{\sigma}\mathbb{E}_{\epsilon\sim\mathcal{N}(0,I)}[f(\theta+\sigma\epsilon)\epsilon]
\end{equation}
In practice, the expectation over Gaussian distribution can be approximated by sampling $n$ noise samples $\{\epsilon_i\}_{i=1,...,n}$ and querying the corresponding function value $\{f_i\}_{i=1,...,n}$:
\begin{equation}
\label{eq:esgradapprox}
\nabla_\theta \hat{f}(\theta)\approx\frac{1}{n\sigma}\sum_{i=1}^{n}f_i\epsilon_i
\end{equation}
As shown in Fig. \ref{fig:alg} (a), one can choose the closed-loop performance (episodic return) as the objective function $f(\theta)$, and directly optimize the parameter $\theta$ by iteratively sampling trajectories using this technique. However, such methods usually suffer from high sample complexity due to the ignorance of the MDP temporal structures.

\subsection{Reinforcement Learning}
\label{sec:prelim.rl}
Reinforcement learning and optimal control problems can be formulated as a \textit{Markov Decision Process} (MDP) defined as $(\mathcal{S}, \mathcal{A}, \mathcal{P}, r)$, where $\mathcal{S}$ is the state space, $\mathcal{A}$ is the action space, $\mathcal{P}: \mathcal{S}\times\mathcal{A}\times\mathcal{S}\to\mathbb{R}$ is the transition probability matrix, $r:\mathcal{S}\times\mathcal{A}\to\mathbb{R}$ is the reward function (or cost function), and $\gamma\in(0,1)$ is the discount factor. The return is defined as the total discounted future reward $G_t=\sum_{i=0}^{\infty}\gamma^{i}r(s_{t+i},a_{t+i})$. State-value function and action-value function is defined as the expected return under policy $\pi$: $V^\pi(s)=\mathbb{E}_{\pi}[G_t|s_t=s]$ and $Q^\pi(s,a)=\mathbb{E}_{\pi}[G_t|s_t=s, a_t=a]$, respectively. 

Consider a deterministic policy $\pi_\theta:\mathcal{S}\to\mathcal{A}$ that directly maps states to actions, which is often the case in control community. The objective of RL is to maximize a cumulative reward (or minimize a cumulative cost) over a sequence of time steps, which can be formally represented as:
\begin{equation}
\begin{aligned}
J(\theta)=\mathbb{E}\left[\sum_{t=0}^{\infty}\gamma^{t}r(s_{t},a_{t})\Big\vert \pi_\theta\right]
=\mathbb{E}_{s\sim d_0}[V^{\pi_\theta}(s)]
\end{aligned}
\end{equation}
where $d_0$ is the initial state distribution.

The deterministic policy gradient (DPG) theorem holds for any differentiable policy $\pi_\theta$ \cite{silver2014deterministic}:
\begin{equation}
\label{eq:dpg}
\nabla_\theta J(\theta)=\mathbb{E}_{s\sim d_{\pi_\theta}}[\nabla_\theta\pi_\theta(s)\nabla_a Q^{\pi_\theta}(s, a)\vert_{a=\pi_\theta(s)}]
\end{equation}
where $d_{\pi_\theta}$ is the discounted state distribution.

Actor-critic methods are among the most popular RL algorithms, which usually introduce two function approximators, one for value function estimation (critic) and another for optimal policy approximation (actor), as shown in Fig. \ref{fig:alg} (b).
The critic is used for bootstrapping in value estimation, which leads to reduced variance and accelerated learning process \cite{sutton2018reinforcement}.

A common choice in deep reinforcement learning (DRL) is to use deep neural networks (DNN) and optimize with first-order optimization techniques. 
However, since such black-box approximators are extremely difficult to interpret or to formally analyze and guarantee safety issues, classic controllers that benefit from rich theories are still preferred in many scenarios.
The problem becomes how to use modern RL techniques to improve the performance of these controllers, given that most of the controllers are not end-to-end differentiable with respect to the design parameters.

\section{PROPOSED FRAMEWORK}
\label{sec:method}

\subsection{Problem Formulation}
\label{sec:method.formulation}
We start by formulating the performance-driven controller tuning problem within the framework of RL. 
Assuming that the users have specified a particular controller structure $\pi_\theta$ and the search space $\Theta$ of parameters, the goal is to find the best parameter $\theta^*$ that minimizes the cumulative cost:
\begin{equation}
\label{eq:problem}
\theta^*=\arg\min_{\theta\in\Theta}J(\theta)
\end{equation}

We will show how to solve (\ref{eq:problem}) using the derivative-free reinforcement learning algorithm ZOAC in section \ref{sec:method.zoac} and restrict and normalize the parameter space to facilitate the tuning process in section \ref{sec:method.paraspace}.

It is worth noting that the state $s\in\mathcal{S}$ and action $a\in\mathcal{A}$ in the MDP do not necessarily have to be identical to that of the real dynamics systems. 
In addition to the current state of dynamics systems, quite a few controllers leverage other information, for example, states and control inputs from several previous steps and reference signals. 
One may always augment them into the state representation to fit the policy form of $a=\pi_\theta(s)$.
Therefore, the state definition is strongly associated with the specific controller structure at hand.
State transitions can be regarded as part of the (unknown) environmental dynamics $\mathcal{P}$ from the perspective of the RL agent.
As for the cost function $r$, it can consist of arbitrary performance indexes that users really care about, such as control performance, power efficiency and constraint violation.
Section \ref{sec:examples} provides the details of two concrete examples from the autonomous driving community.

\subsection{Zeroth-Order Actor-Critic}
\label{sec:method.zoac}

As described in section \ref{sec:prelim.zog}, in evolutionary methods like ES, the policy is perturbed in parameter space at the beginning of an episode and remains unchanged throughout the trajectories. 
If a large number of random directions $n$ is evaluated, the sample complexity will increase significantly.
However, since the zeroth-order gradient is estimated as the weighted sum of several random directions, it exhibit excessively high variance when $n$ is small, which may greatly harm the performance.
Therefore, it is essential to trade-off the contradictory between sample efficiency and variance.

To encourage sufficient exploration and low variance while maintaining high sample efficiency, here we consider perturbating the policy at every timestep, resulting in a stochastic exploration policy $\beta=\pi_{\theta+\sigma\epsilon}$ perturbed from the deterministic policy $\pi_\theta$, where $\sigma$ is the standard deviation and $\epsilon\sim\mathcal{N}(0,I)$ is Gaussian parameter noise sampled i.i.d. at every timestep. Our objective is to optimize the cumulative cost and we can derive Theorem 1 under this setting.


\textbf{Theorem 1.} \textit{For MDP that satisfies regularity conditions in the Appendix A, the zeroth-order policy gradient of the Gaussian smoothed objective $J(\theta)=\mathbb{E}_{s\sim d_0}[V^\beta(s)]$ is:}
\begin{align}
\label{eq:zoac.orig}
\nabla_\theta J(\theta)=\frac{1}{\sigma}\mathbb{E}_{s\sim d_{\beta}}\mathbb{E}_{\epsilon\sim\mathcal{N}(0,I)}[Q^\beta(s,\pi_{\theta+\sigma\epsilon}(s))\epsilon]
\end{align}

where $d_{\beta}$ is the discounted state distribution and can be approximated with on-policy samples.
The proof adopts a similar scheme to \cite{silver2014deterministic} and is provided in Appendix A.
The derivation can be further extended to the case where the behavior policy $\beta$ runs forward $N$ timesteps with each sampled noise $\epsilon$ instead of one timestep only.

From this foundation, we are able to integrate derivative-free optimization policy improvement into the RL framework as shown in Fig. \ref{fig:alg} (b).
A common technique in on-policy actor-critic methods to speed up learning is replacing the action-value function $Q(s,a)$ in the policy gradient with a proper form of advantage function. 
In the practical tuning algorithm, we also use generalized advantage estimation (GAE) \cite{schulman2015high} to construct the advantage function, which has been shown to effectively strike a good balance between bias and variance of the value target. 

\begin{algorithm}[h]
\caption{Controller Tuning via ZOAC}
\begin{algorithmic}[1]
\label{code}
\STATE \textbf{Initialize:} controller parameters $\theta \in \Theta$ ( $\theta_{m}\in\Theta_m$ correspondingly), critic network parameters $w$
\FOR{each iteration}
\FOR{each worker $i=1,2,...,n$ (parallelizable)}
\FOR{$j=0,1,...,H-1$}
\STATE Sample $\epsilon_{i,j}\sim \mathcal{N}(0,I)$
\STATE Run perturbed controller $\pi_{i,j}$ (w/ parameter $\theta_{i,j}$ that mapped from $\theta_m+\sigma\epsilon_{i,j}$) for $N$ time steps\label{code:1}
\STATE Compute advantage $\hat{A}_{i,j}$ according to (\ref{eq:zoac.advest})
\ENDFOR
\STATE Compute the state-value target $\hat{G}$ for each state $s$ according to (\ref{eq:zoac.tarest})
\ENDFOR
\STATE Collect $(s,\hat{G})$ and $(\epsilon, \hat{A})$ for critic and actor update
\STATE Update $w$ with batch size $L$ through SGD by minimizing (\ref{eq:zoac.pev}) for $M$ epoches
\STATE Update $\theta_m$ along the zeroth-order gradient direction estimated in (\ref{eq:zoac.pim}) and update $\theta$ thereafter\label{code:2}
\ENDFOR
\STATE \textbf{Output:} Optimal controller parameters $\theta^*$
\end{algorithmic}
\end{algorithm}

The pseudocode of the proposed controller tuning method is summarized in Algorithm \ref{code}.
In each iteration, the following two processes update the critic and actor with on-policy samples collected under $\beta$ respectively.

\subsubsection{Policy Evaluation (PEV)}
\label{sec:method.zoac.pev}
In a trajectory with length $T$, the value target $\hat{G}$ for each state $s$ in the trajectory can be calculated:
\begin{equation}
\label{eq:zoac.tarest}
\hat{G}=V_w(s)+\sum_{k=0}^{T-1}(\gamma\lambda)^k[r_{k}+\gamma V_w(s_{k+1})-V_w(s_{k})]
\end{equation}
where $0<\lambda<1$ is a hyperparameter of GAE.
In practice, the critic is a neural network with parameter $w$ and updated through epochs of stochastic gradient descent by minimizing:
\begin{equation}
\label{eq:zoac.pev}
J_{\mathrm{critic}}(w)=\mathbb{E}_{(s,\hat{G})}\left[\frac{1}{2}\left(V_w(s)-\hat{G}\right)^2\right]
\end{equation}

\subsubsection{Policy Improvement (PIM)}
In a trajectory segment with length $N$ that is unrolled under controller $\pi_{i,j}$, the advantage function is calculated as:
\begin{equation}
\begin{aligned}
\label{eq:zoac.advest}
\hat{A}_{i,j}=\sum_{k=0}^{N-1}(\gamma\lambda)^k[r_k+\gamma V_w(s_{k+1})-V_w(s_{k})]
\end{aligned}
\end{equation}
where $\lambda$ is the same as in (\ref{eq:zoac.tarest}).
The zeroth-order gradient can be then estimated as the weighted sum of the sampled random directions:
\begin{equation}
\label{eq:zoac.pim}
\nabla_\theta J_{\mathrm{actor}}(\theta)\approx\frac{1}{nH\sigma}\sum_{i=1}^{n}\sum_{j=0}^{H-1}\hat{A}_{i,j}\epsilon_{i,j}
\end{equation}


The main advantage of the proposed algorithm is twofold: 
For one thing, it conducts policy improvement by following the estimated zeroth-order policy gradient in the actor-critic framework (\ref{eq:zoac.pim}).
The derivative-free nature ensures its high versatility, i.e., it can be applied seamlessly to arbitrary parameterized controllers in theory, no matter differentiable or not.
Moreover, the highly noisy Monte Carlo return used in standard ES gradient estimator (\ref{eq:esgradapprox}) is replaced by the advantage function (\ref{eq:zoac.pim}), which efficiently utilize the sequential nature.
For another, it collects experiences with timesteps-wise exploration in parameter space, which is neither similar to DFO \cite{salimans2017evolution} nor DPG-based RL \cite{silver2014deterministic}. Instead, the former remains the same parameter unchanged throughout the whole trajectory, while the latter usually explores with unstructured noises added to the control input. 
The state-dependent and temporally-extended exploration leads to high sample efficiency and strong optimization ability.
The asymptotic convergence property of such two time-scale actor-critic methods under certain assumptions has been well studied \cite{konda1999actor}. However, for various practical algorithms, especially when neural approximators are involved, the convergence and optimality analysis is still largely open in the community and will be our future work.

\subsection{Restricting and Normalizing the Parameter Space}
\label{sec:method.paraspace}
Parameters in the designed controller are usually restricted according to the specific task at hand, prior knowledge available, etc \cite{piga2019performance}.
For example, in the dynamic models that are derived from first principles, each parameter has its corresponding physical meaning (e.g., mass, length, rotation inertia) and reasonable interval. 
Existing auto-tuning methods like Bayesian optimization usually allow users to set bounds on the search space. 
We also restrict the search space to facilitate the tuning process in this work. 
Formally, $\Theta \subseteq \mathbb{R}^d$ is a box-constrained search space determined by the lower and upper bounds of each parameter, where $d$ is the number of design parameters.

Since the exploration during the learning process relies on local random search by adding isotropic Gaussian noise in the parameter space, it is critical to ensure the behavioral diversity of the perturbed policies in each iteration.
In the controller tuning problem, we achieve this goal by searching in a mapped parameter space $\Theta_m = [-1, 1]^d$, in which each dimension is a one-to-one mapping of the original search space $\Theta$.
Linear mapping (e.g., model parameters, control gains) and logarithmic mapping (e.g., diagonal positive definite weight matrices) are used in this work.
The minor modification is revealed in Line \ref{code:1} \& \ref{code:2} in Algorithm \ref{code}.

\section{NUMERICAL EXPERIMENTS}
\label{sec:examples}

\begin{figure}[tb]
\centering
\subfigure[ACC]{\includegraphics[width=0.3\linewidth, keepaspectratio=true,trim=100 110 500 120,clip]{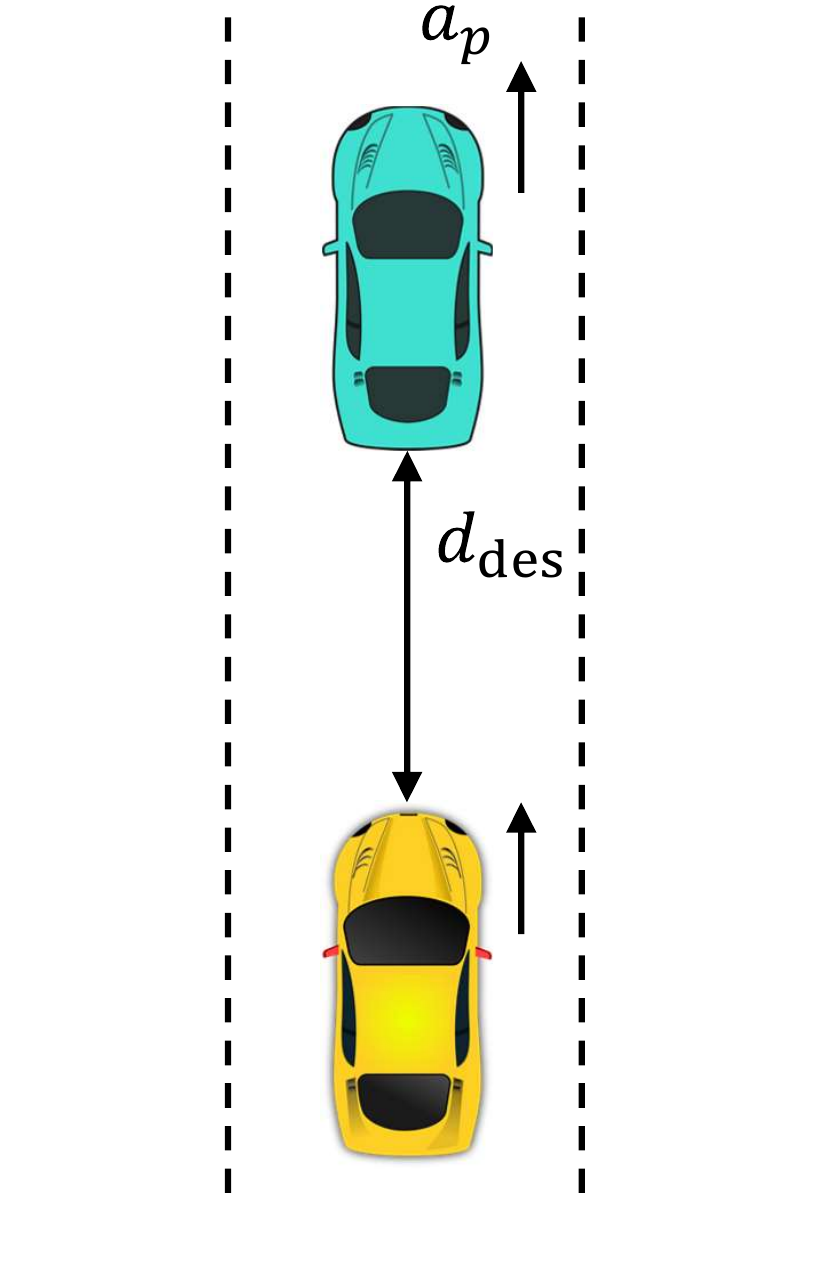}}
\hspace{0.3in}
\subfigure[Trajectory tacking \cite{ge2021numerically}]{\includegraphics[width=0.36\linewidth, keepaspectratio=true,trim=0 0 50 120,clip]{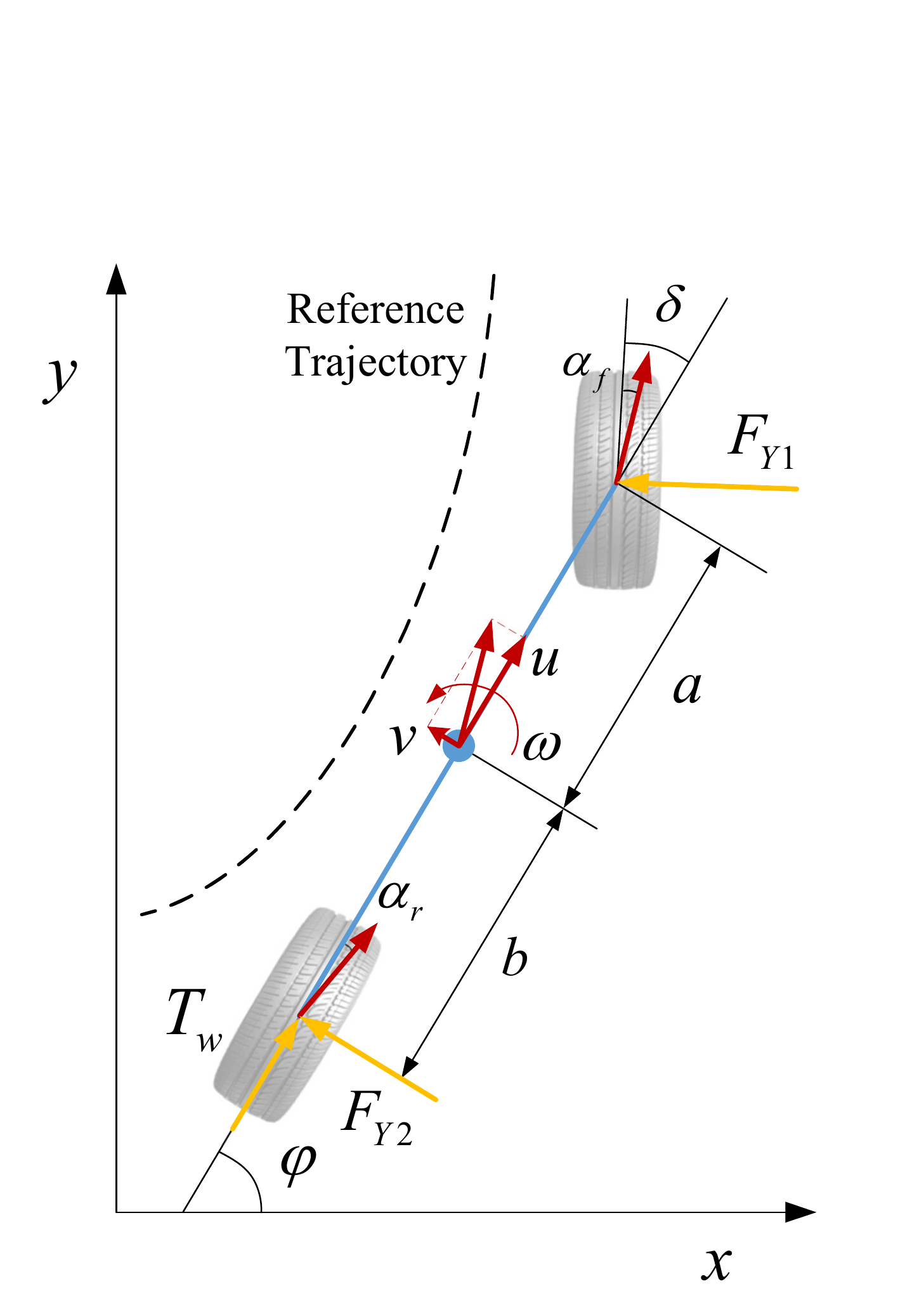}}
\vspace{-0.05in}
\caption{Schematics of two tasks.}
\vspace{-0.1in}
\label{fig:tasks}
\end{figure}

We apply the proposed controller tuning framework to two common autonomous driving tasks: adaptive cruise control (ACC) \cite{li2010model} and trajectory tracking \cite{ge2021numerically}, as shown in Fig. \ref{fig:tasks}.
We compare our methods with two popular baselines: \textit{Bayesian optimization} (BO) and \textit{covariance matrix adaptation evolution strategies} (CMA-ES).
We use their off-the-shelf versions provided in NeverGrad \cite{nevergrad}, an open-source gradient-free optimization platform.
We follow the recommended settings for baselines and summarize the hyperparameters of ZOAC in Table \ref{tab:hyper}.
\subsection{Tuning PID for ACC}
\label{sec:examples.acc}

\begin{figure*}[htb]
\centering
\subfigure[Cumulative cost in PID-ACC task]{\includegraphics[width=0.28\linewidth, keepaspectratio=true,trim=0 10 0 10,clip]{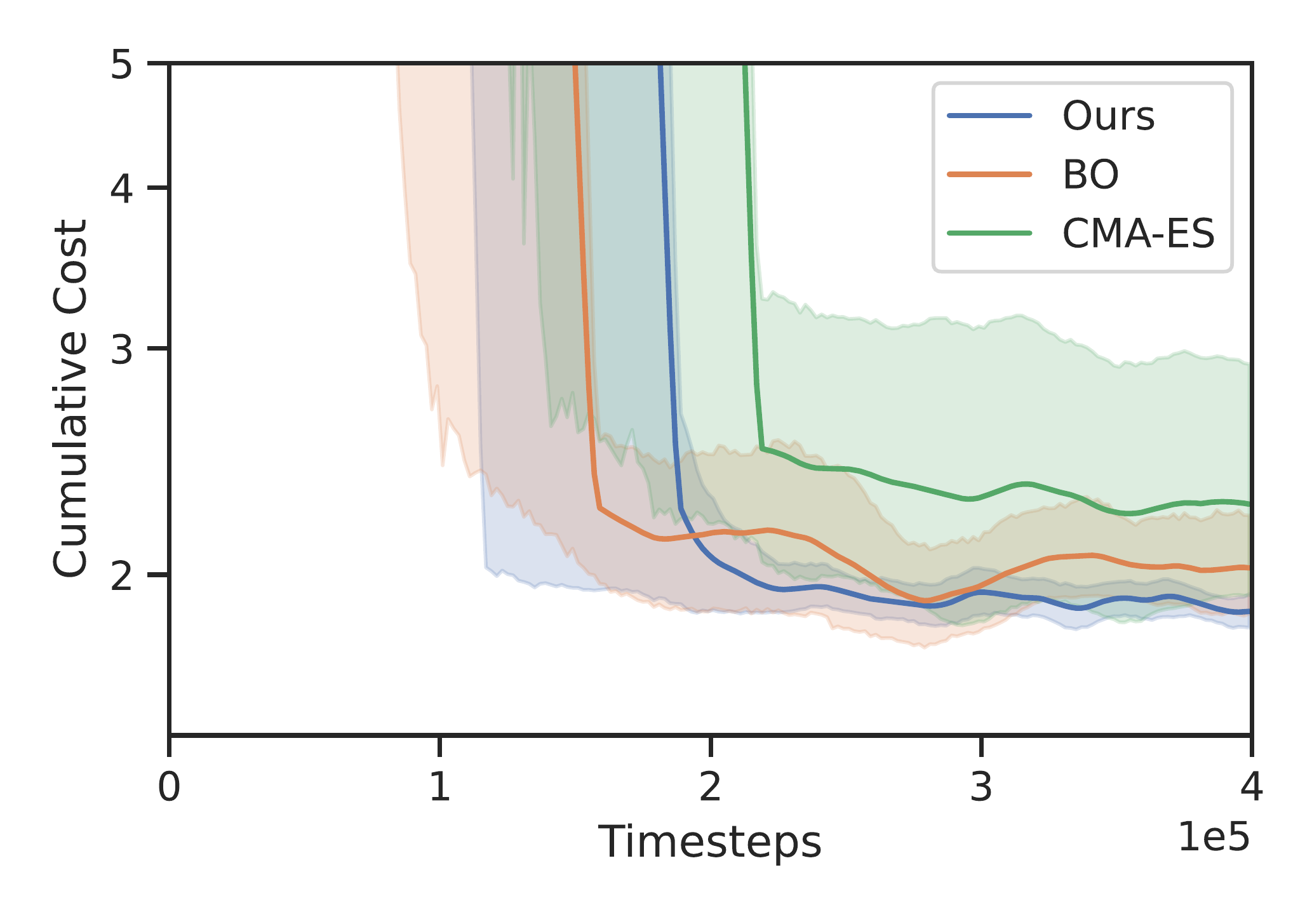}}
\subfigure[Cumulative cost in MPC-Tracking task]{\includegraphics[width=0.28\linewidth, keepaspectratio=true,trim=0 10 0 10,clip]{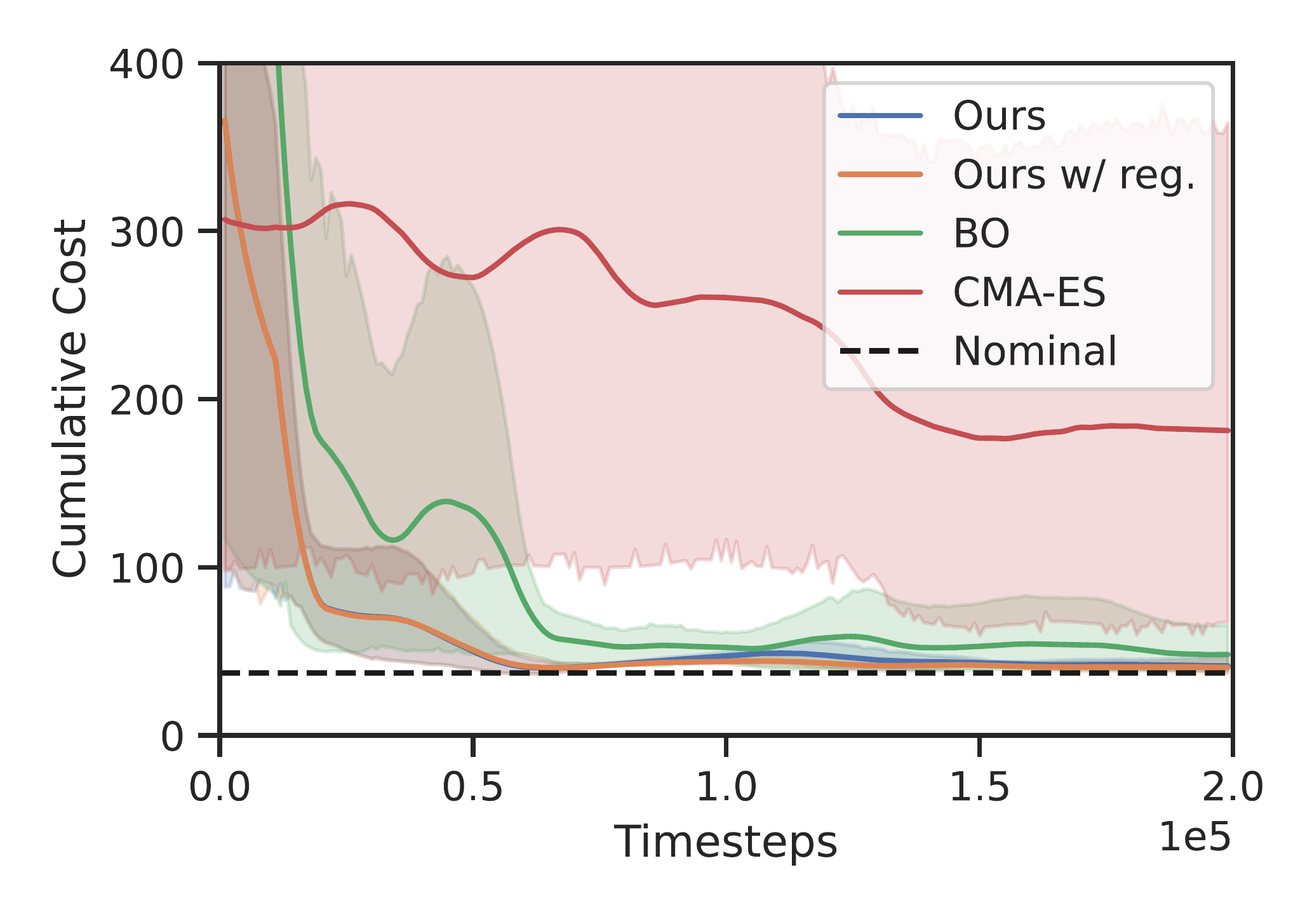}}
\subfigure[Model accuracy in MPC-Tracking task]{\includegraphics[width=0.28\linewidth, keepaspectratio=true,trim=0 10 0 10,clip]{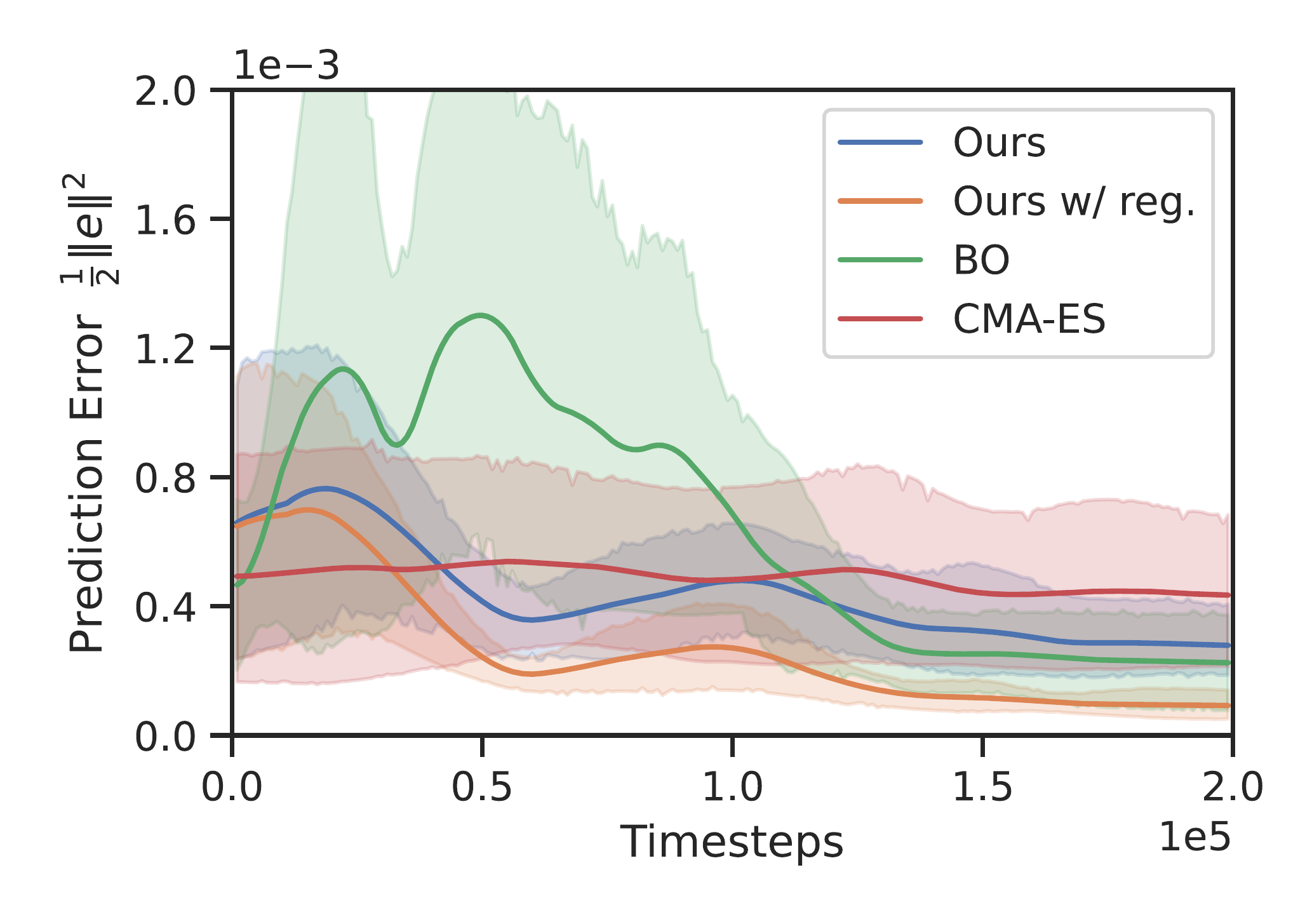}}
\vspace{-0.05in}
\caption{Learning curves: (a) PID-ACC task; (b)-(c) MPC-Tracking task. The solid lines correspond to the mean and the shaded regions to the 95\% confidence interval over multiple trials. All curves are smoothed uniformly for visual clarity.}
\vspace{-0.1in}
\label{fig:result.curve}
\end{figure*}

Adaptive cruise control (ACC) is a widely-used advanced driver-assistance system, which automatically adjusts the speed of ego vehicle to maintain a desired distance from the preceding vehicle \cite{li2010model}, as in Fig. \ref{fig:tasks} (a). 
A hierarchical control structure is often used in ACC system design, where the upper layer outputs the desired acceleration $a_{\mathrm{des}}$ and the lower layer controls the vehicle to track this acceleration. 

Here we mainly focus on the tuning problem of the upper controller. 
Typically, the lower layer and the vehicle together can be viewed as a first-order system $a_f=\frac{K_L}{T_L s+1}a_{\mathrm{des}}$, where $a_f$ is the acceleration of ego vehicle, $K_L=1.0$ is the system gain, and $T_L=0.45$ is the time constant.
We employ the constant time headway policy described as $d_{\mathrm{des}}=\tau_h v_f +d_0$, where $d_{\mathrm{des}}$ is the desired inter-vehicle distance, $\tau_h=2.5$s is the nominal time headway, and $d_0=5$m is the stopping distance for typical drivers. 
The car-following system can then be described as:

\begin{equation}
\begin{aligned}
&\qquad\qquad\qquad\quad \dot{x}=A x + B u + D w \\
&A=\left[{\small{\begin{array}{ccc}
0 & 1 & -\tau_h \\
0 & 0 & -1 \\
0 & 0 & -1/{T_L}
\end{array}}}\right],
B=\left[{\small{\begin{array}{c}
0\\
0\\
K_L/T_L
\end{array}}}\right],
D=\left[{\small{\begin{array}{c}
0\\
1\\
0
\end{array}}}\right]
\end{aligned}
\end{equation}

\begin{figure}[ht]
\centerline{\includegraphics[width=0.99\linewidth, keepaspectratio=true,trim=10 140 10 10,clip]{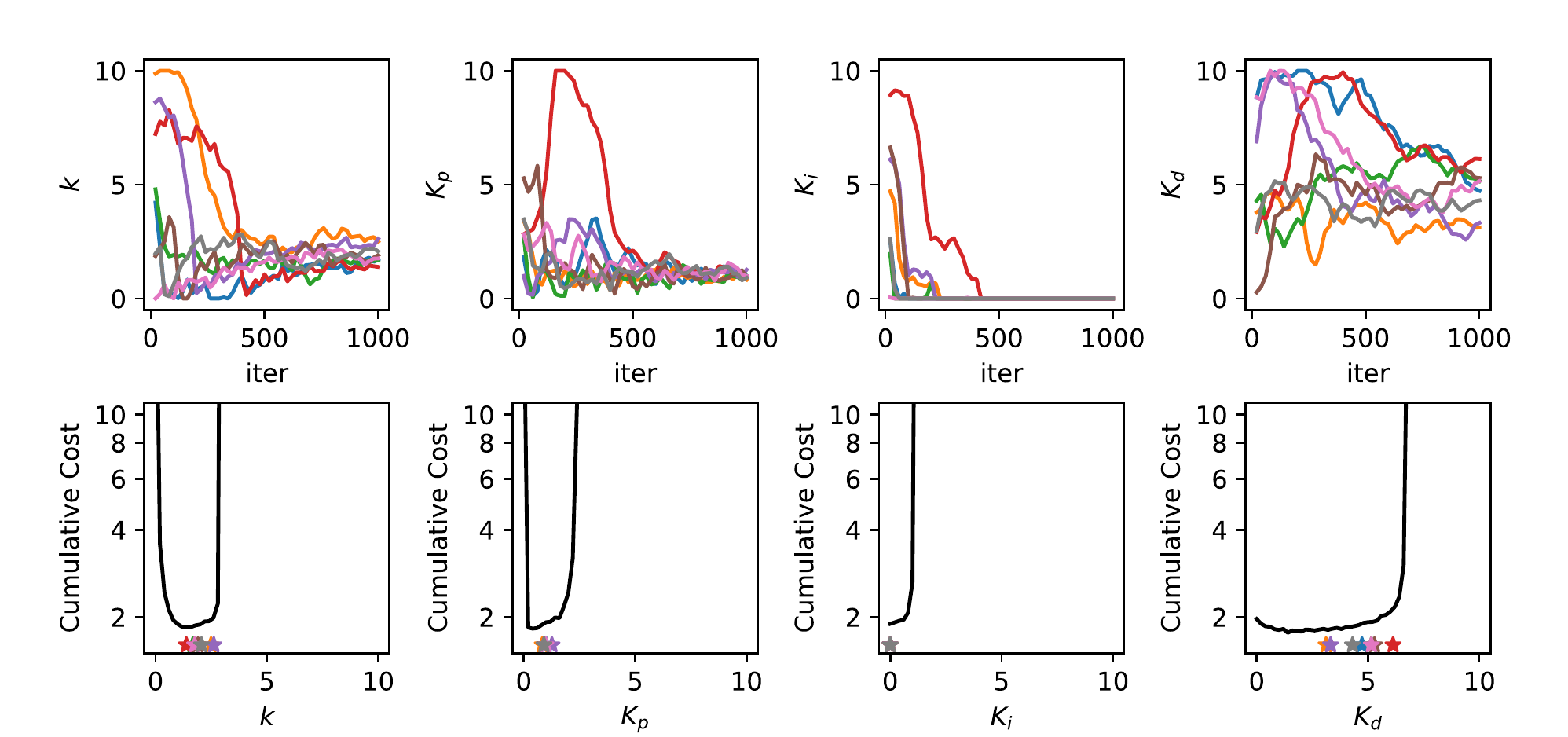}}
\caption{Parameter evolution process and approximate 1-D landscape.}
\label{fig:acc.conv_landscape}
\vspace{-0.1in}
\end{figure}


The system state vector is $x =[\Delta d \quad \Delta v \quad a_f]^{T}$ , where $\Delta d=d-d_{\mathrm{des}}$ is the clearance error, $\Delta v=v_p-v_f$ is the speed error with $v_p$ denoting the speed of preceding vehicle and $v_f$ denoting the speed of ego vehicle, and $a_f$ is the acceleration of ego vehicle.
The control input $u=a_{\mathrm{des}}$ is the desired acceleration of ego car, and the disturbance of the system $w=a_{\mathrm{des}}$ is the acceleration of preceding vehicle.

Consider to use a simple PID controller:

\begin{equation}
\begin{aligned}
&e=k \Delta d+\Delta v \\
&u=K_p e+K_i\int{e\mathrm{d}t}+K_d\frac{\mathrm{d}e}{\mathrm{d}t}
\end{aligned}
\end{equation}

where $k$ is the coefficient that trades off between clearance error and speed error, $K_p, K_i, K_d$ are proportional gain, integral gain and differential gain, respectively.
In practice, we use an incremental version of PID and the control input $u$ is limited within the range from $-1.5\mathrm{m/s^2}$ to $0.6\mathrm{m/s^2}$:

\begin{equation}
\begin{aligned}
\label{eq:pidcontrol}
\Delta u_t&=K_p(e_t-e_{t-1})+K_i e_t+K_d(e_t-2e_{t-1}+e_{t-2})\\
u_t&=u_{t-1}+\Delta u_t
\end{aligned}
\end{equation}


As illustrated in section \ref{sec:method.formulation}, we can construct an MDP trivially according to (\ref{eq:pidcontrol}).
The 1-dimensional action is the control input increment, i.e., $a=\Delta u$.
The 10-dimensional state is augmented with states and control inputs from several previous steps, i.e., $s=(x_t, x_{t-1}, x_{t-2}, u_{t-1})$.
The cost $r$ is borrowed from \cite{li2010model}, which considers several important aspects, including tracking error, fuel efficiency, riding comfort, and similarity to professional truck drivers:

\begin{equation}
\begin{aligned}
r=0.1\Delta v^2&+0.06\Delta d^2+ a_{\mathrm{des}}^2+0.1\dot{a}_{\mathrm{des}}^2\\&+0.5(0.25\Delta v+0.02\Delta d-a_f)^2
\end{aligned}
\end{equation}

There are four design parameters $\theta=(k, K_p, K_i, K_d)$ that need to be optimized within $\Theta=[0,10]^4$. For a fair comparison, we use the mapped search space $\Theta_m$ described in section \ref{sec:method.paraspace} for all methods. 
In our experiments, the maximum episode length is set as 1000, with a simulation time step of 0.1s. 
The acceleration of preceding vehicle $a_p$ is sampled from $\mathcal{N}(0, 0.05)$ every 3s during training.  
When the error is too large ($\vert \Delta d\vert>5$ m or $\vert \Delta v\vert>1$ m/s), an episode ends and an additional penalty $r_p=1000$ is incurred.

\begin{figure}[t]
\vspace{0.1in}
\centerline{\includegraphics[width=0.8\linewidth, keepaspectratio=true,trim=20 30 20 20,clip]{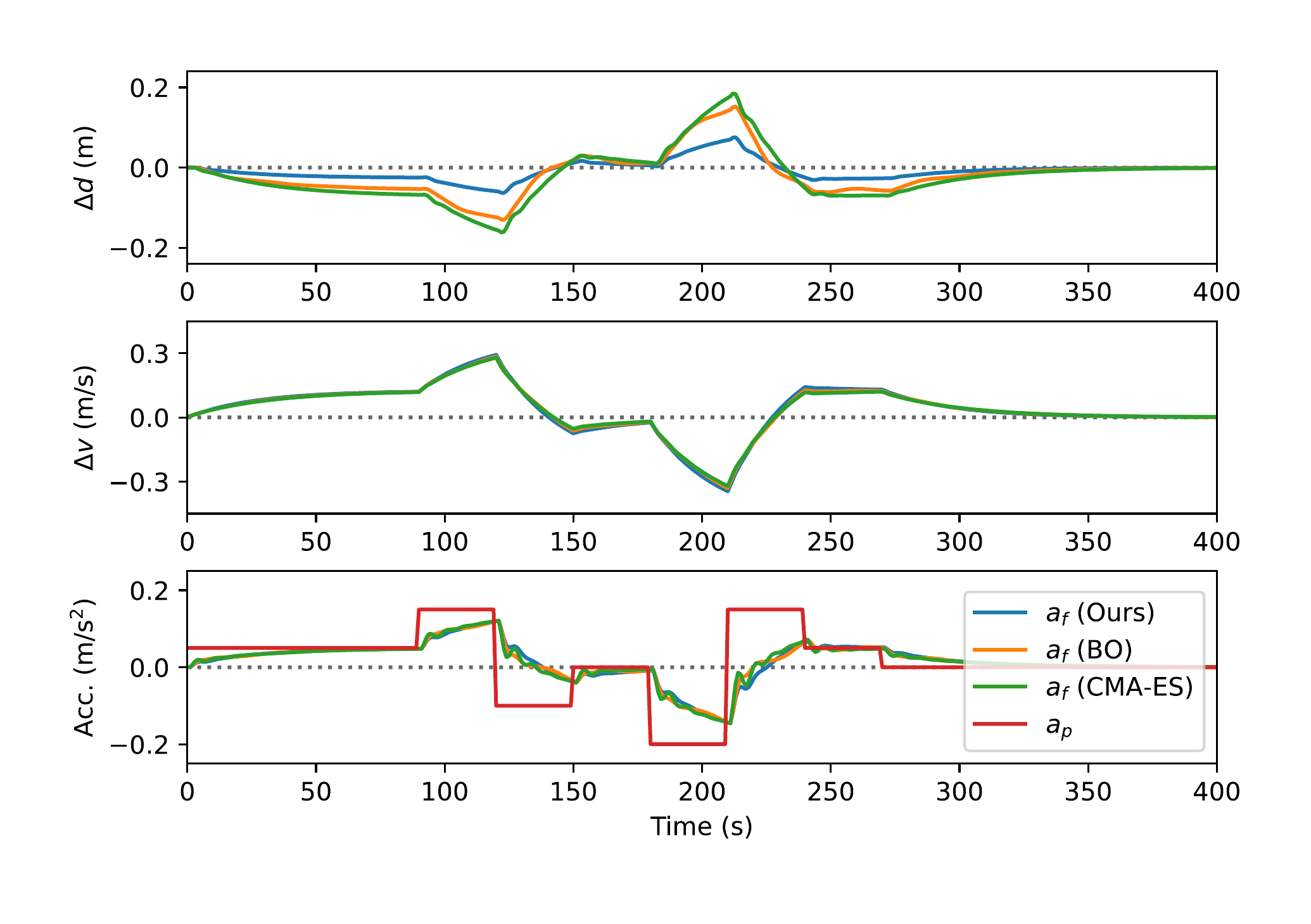}}
\caption{A typical trajectory of the tuned PID controller.}
\vspace{-0.1in}
\label{fig:acc.closeloop}
\end{figure}

Fig. \ref{fig:result.curve} (a) shows the learning curves on this PID-ACC task over eight independent trials using different random seeds, where our method outperforms baselines in final performance.
To empirically visualize the convergence and optimality of our method, we plot the parameter evolution process during eight trials and also the 1-dimensional landscape (approximate via Monte Carlo estimation) in Fig. \ref{fig:acc.conv_landscape}.
Besides, a typical closed-loop trajectories of the PID controller tuned by different methods is plotted in Fig. \ref{fig:acc.closeloop}.
Results show that all parameters converge quickly to the optima starting from \textit{random} initial values, and the resulting controllers achieve satisfying performance, which strongly supports the effectiveness of our method.

\subsection{Tuning MPC for Trajectory Tracking}
\label{sec:examples.track}
Trajectory tracking is an essential module in the autonomous driving pipeline. As shown in Fig. \ref{fig:tasks} (b), it is responsible to track the reference trajectory from the upper decision-making module by regulating the vehicle's accelerator (or brake) and steering wheels.

Model predictive control (MPC) is widely used as an integrated longitudinal and lateral controller for autonomous vehicles due to its ability to explicitly optimize a surrogate cost function that considers multiple objectives within a finite predictive horizon. 
A typical MPC controller for trajectory tracking optimizes the following problem repeatedly online:

\begin{equation}
\begin{aligned}
&\min_{X,U} \sum_{k=0}^{N_p-1} \gamma^k l_s(X(k),X_r(k),U(k))+\gamma^{N_p}l_t(X(N_p),X_r(N_p))\\
&\text{s.t.} \qquad\qquad\qquad X(0)=X_t&\\
&\qquad X(k+1)=F(X(k),U(k)),\quad k=0,1,\cdots,N_p-1\\
&\qquad X_{\mathrm{min}}\leq X(k)\leq X_{\mathrm{max}},\quad k=0,1,\cdots,N_p\\
&\qquad U_{\mathrm{min}}\leq U(k)\leq U_{\mathrm{max}},\quad k=0,1,\cdots,N_p-1
\end{aligned}
\end{equation}

where $X_t$ is the state of the vehicle at time $t$, $X(k)$ is the predicted states in the virtual horizon, and $N_p=25$ is the predictive horizon. 
The problem is repeatedly solved at every timestep and only the first action $U(0)$ in the sequence is applied as $U_t$.
The six state variables are respectively horizontal position $x$, vertical position $y$, yaw angle $\varphi$, longitudinal velocity $u$, lateral velocity $v$ and yaw rate $\omega$. 
The control input $U$ includes front wheel steering angle $\delta$ and acceleration command $a$. In practice, we constrain $\delta$ within $\left[-\frac{2\pi}{15}, \frac{2\pi}{15}\right]$ rad and $a$ within $[-3,3]$ m/s$^2$ but we do not explicitly constrain the states.

We use a numerically stable dynamic bicycle model \cite{ge2021numerically} as the forward predictive model $F$:

\begin{equation}
    \left[\begin{array}{c}
    x_{k+1} \\
    y_{k+1} \\
    \varphi_{k+1} \\
    u_{k+1} \\
    v_{k+1} \\
    \omega_{k+1}
    \end{array}\right]=\left[\begin{array}{c}
    x_{k}+T_{s}\left(u_{k} \cos \varphi_{k}-v_{k} \sin \varphi_{k}\right) \\
    y_{k}+T_{s}\left(v_{k} \cos \varphi_{k}+u_{k} \sin \varphi_{k}\right) \\
    \varphi_{k}+T_{s} \omega_{k} \\
    u_{k}+T_{s} a_{k} \\
    \frac{m u_{k} v_{k}+T_{s}\left(l_{f}k_{f}-l_{r} k_{r}\right) \omega_{k}-T_{s} k_{f} \delta_{k} u_{k}-T_{s} m u_{k}^{2} \omega_{k}}{m u_{k}-T_{s}\left(k_{f}+k_{r}\right)} \\
    \frac{I_{z} u_{k} \omega_{k}+T_{s}\left(l_{f} k_{f}-l_{r} k_{r}\right) v_{k}-T_{s} l_{f} k_{f} \delta_{k} u_{k}}{I_{z} u_{k}-T_{s}\left(l_{f}^{2} k_{f}+l_{r}^{2} k_{r}\right)}
    \end{array}\right]
    \label{eq:track.model}
\end{equation}

\begin{table}[b]
    \caption{Parameters in the dynamic bicycle model}
    \label{tab:paras}
    \vspace{-0.05in}
    \begin{center}
    \begin{tabular}{cccc}
    \hline
    \textbf{Parameter}&\textbf{Description}&\textbf{Value}&\textbf{Interval}\\
    \hline
{$I_{z}$ (kg$\cdot$m$^2$)} & yaw inertia of vehicle body & 1536.7 & [1e3, 2e3]\\
    $k_f$ (N/rad) & front axle sideslip stiffness & -128916 & [-16e4,-8e4] \\
    $k_r$ (N/rad) & rear axle sideslip stiffness & -85944 & [-16e4, -8e4]\\
    $l_f$ (m) & distance from C.G. to front axle & 1.06 & [0.8, 2.2]\\
    $l_r$ (m) & distance from C.G. to rear axle & 1.85 & [0.8, 2.2]\\
    $m$ (kg) & mass of the vehicle & 1412 & [1e3, 2e3]\\
    \hline
    \end{tabular}
    \end{center}
\end{table}

The physical meanings of the parameters involved are listed in Table \ref{tab:paras}, and $T_s=0.1$s is the step length.
We assume that only a rough interval is known for each model parameter but not the exact value and hence needs to be tuned. 

The quadratic function is a common choice for stage cost $l_s$ and terminal cost $l_t$ for real-time performance:

\begin{equation}
\begin{aligned}
&l_s(X(k),X_r(k),U(k))\\
&=(X(k)-X_r(k))^T Q_{s}(X(k)-X_r(k))+U(k)^TRU(k)\\
&\qquad\qquad\qquad\qquad\qquad\qquad\qquad k=0, 1, \cdots, N_p-1\\
&l_t(X(N_p),X_r(N_p))\\
&=(X(N_p)-X_r(N_p))^T Q_{t}(X(N_p)-X_r(N_p))
\end{aligned}
\end{equation}

\begin{figure}[bp]
\centerline{\includegraphics[width=0.8\linewidth, keepaspectratio=true,trim=30 40 30 30,clip]{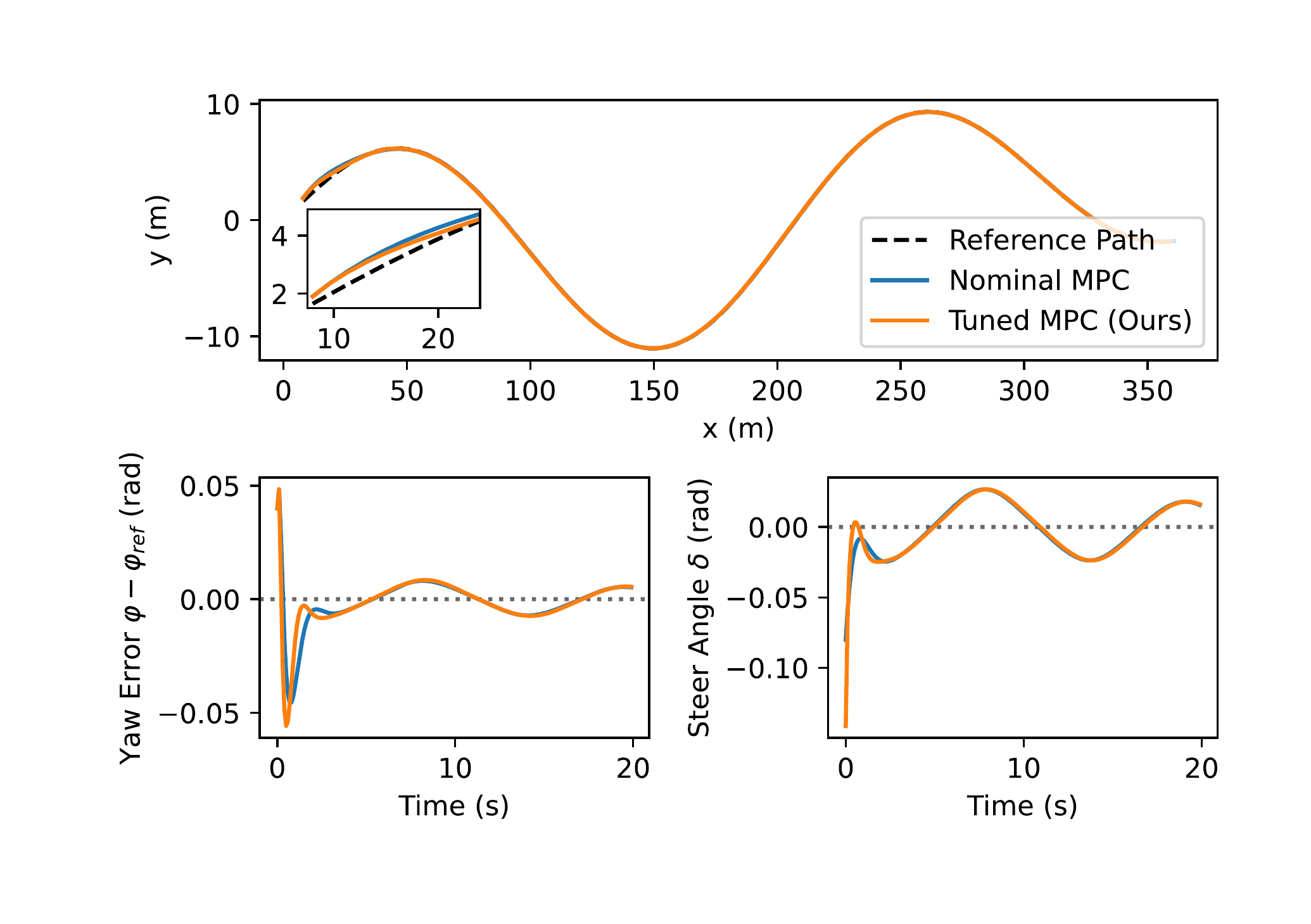}}
\caption{Trajectory tracking performance of the tuned MPC controller.}
\label{fig:track.closeloop}
\end{figure}

where $Q_{\diamondsuit}=\mathrm{diag}(w_\diamondsuit^x,w_\diamondsuit^y,w_\diamondsuit^\varphi,w_\diamondsuit^u,w_\diamondsuit^v,w_\diamondsuit^\omega), \diamondsuit\in \{s, t\}$ and $R=\mathrm{diag}(w^\delta,w^a)$ are diagonal positive-definite weighting matrices that need proper tuning to approximate the objective of real concern. 
Since scaling all parameters simultaneously does not change the optimal solution, we fix $w_s^u$ as $10^{-2}$ in our experiments and optimize the remainder within the search space $[10^{-6}, 10^{2}]^{13}$.

To sum up, there are altogether 19 parameters that need tuning in this task, including those in the predictive model $F$ and the weighting matrices $Q_s$, $Q_t$ and $R$. 
As illustrated in section \ref{sec:method.paraspace}, for all methods, the former ones are searched in a linearly mapped space while the latter ones are searched in a logarithmically mapped space.

As for the MDP construction, the action is identical to the control input, while the state includes $X_t$ at the current timestep and all $N_p$ reference signals $X_r(k)$ in the predictive horizon. 
The reference signals given by the upper decision-making module are reasonably assumed to satisfy the Markov property.
The cost signal $r$ mainly considers tracking error and riding comfort but is still constructed as a quadratic function with an additional penalty $r_p$ for simplicity, which in fact, could includes more complex performance indexes (e.g., fuel consumption):

\begin{equation}
\begin{aligned}
r=&(u-u_r)^2+4(y-y_r)^2+10(\varphi-\varphi_r)^2\\
&+2\omega^2+500\delta^2+5a^2
\end{aligned}
\end{equation}

The maximum episode length is 200, with a simulation time step of 0.1s. An episode ends earlier whenever the tracking error is too large ($\vert y-y_r\vert>4$ m or $\vert \varphi-\varphi_r\vert>\pi/4$ rad or $\vert u-u_r\vert>4$ m/s) and $r_p=1000$ is incurred.

Fig. \ref{fig:result.curve} (b)-(c) show the learning curves of cumulative cost and one-step prediction error over five independent trials using different random seeds, respectively. 
Table \ref{tab:finalp} compares the final performance of the controllers tuned with different methods. 
A typical trajectory of the tuned MPC controller is also plotted in Fig. \ref{fig:track.closeloop}. 
On the whole, our tuning method performs well in this task, while CMA-ES seems get stuck easily and BO suffers from larger variance.
The required tuning time of baselines (all methods run on Intel i5-10400F 2.9GHz) is also much longer than ours, since they suffer significantly increased computational burden in high-dimensional cases.
Moreover, the tuned controller achieves very similar performance to the \textit{nominal} MPC controller, where vehicle dynamics and cost function are known a priori.

We further conduct an ablation study by augmenting the original cost signal with a regularization term indicating the model's accuracy, i.e., $r'=r+\frac{\zeta}{2}\Vert F(X_t,U_t) - X_{t+1}\Vert_2^2$, where $\zeta$ is the regularization coefficient.
From the results presented in Fig. \ref{fig:result.curve} and Table \ref{tab:finalp}, it can be observed that this regularization term indeed improves the accuracy of the predictive model, but has a minor effect on the final performance. 
A possible explanation is that replanning at every timestep resolves the model inaccuracy to a great extent. 
Hence, using merely closed-loop performance does not provide enough encouragement to improve the model used in MPC. 
The results recover the rationale of I4C, which claims that identifying a globally correct model is often unnecessary and that a good enough model can be learned using the closed-loop performance under a given task.

\begin{table}[b]
\centering
\vspace{-0.1in}
    \caption{Performance comparison on MPC-Tracking task}
    \label{hyperpara}
    \vspace{-0.05in}
    \begin{tabular}{c c c c}
        \hline
        \textbf{Methods} & \textbf{Cumulative cost} & \textbf{Prediction error} & \textbf{Tuning time (s)}\\
          \hline
        \textbf{Ours} & 41.7$\pm$3.8 & (2.9$\pm$1.4)$\times$1e-4 & (6.5$\pm$0.2)$\times$1e3\\
        \textbf{Ours (w/reg)} & 40.5$\pm$3.2 & (9.6$\pm$4.5)$\times$1e-5 & (7.1$\pm$0.1)$\times$1e3\\
        BO & 51.3$\pm$22.9 & (2.3$\pm$1.7)$\times$1e-4 & (2.7$\pm$0.1)$\times$1e4\\
        CMA-ES & 185.0$\pm$184.6 & (4.4$\pm$2.9)$\times$1e-4 & (2.6$\pm$0.1)$\times$1e4\\
        \hline
        \textit{Nominal} & 37.16 & 0.0 & /\\
        \hline
    \end{tabular}
    \label{tab:finalp}
\end{table}

\section{CONCLUSIONS}
\label{sec:conclusion}

This paper proposes a novel derivative-free reinforcement learning framework for performance-driven controller auto-tuning.
The framework formulates controller tuning as a reinforcement learning problem and applies the zeroth-order actor-critic algorithm to optimize the controller parameters within the user-defined search space.
The proposed tuning algorithm adopts the advanced actor-critic RL architecture but still performs derivative-free policy updates, which leads to higher efficiency compared to traditional derivative-free methods while maintaining versatility for various types of controllers.
Numerical experiments on two concrete examples from autonomous driving strongly demonstrate the framework's potential for controller tuning.

In the future, we will devote ourselves to further theoretical analysis and empirical validation on real-world systems of the proposed framework.
The combination with other advanced RL techniques will also be explored, including sample reuse and safe exploration.
Besides, we may consider to extend the framework to suit different tuning requirements, e.g., a discrete optimization variant when the set of candidate controllers is finite in number.

\begin{table}[h]
    \caption{Hyperparameters of ZOAC}
    \label{tab:hyper}
    \vspace{-0.08in}
    \begin{center}
    \begin{tabular}{lc}
    \hline
    \textbf{Hyperparameter}&\textbf{Value}\\
    \hline
    \emph{Shared} & \\
    Number of workers $n$ & 10\\
    Discount factor $\gamma$ & 0.99 \\
    GAE $\lambda$ & 0.95\\
    Critic update epoch $M$ & 10\\
    Minibatch size $L$ & 128\\
    Critic network & (256, 256) w/ tanh\\
    Critic optimizer & Adam($\beta_1$ = 0.9, $\beta_2$ = 0.999)\\
    Critic learning rate & $\alpha_\text{critic} = \text{5e-4}$\\
    Actor optimizer & Adam($\beta_1$ = 0.9, $\beta_2$ = 0.999)\\
    \hline
    \emph{PID-ACC}\\
    Segment length $N$ & 10\\
    Train frequency $H$ & 4\\
    Noise standard deviation $\sigma$ & 0.08\\
    Actor learning rate & Linear annealing 3e-2$\to$1e-2\\
    \hline
    \emph{MPC-Tracking}\\
    Segment length $N$ & 20\\
    Train frequency $H$ & 5\\
    Noise standard deviation $\sigma$ & 0.1\\
    Actor learning rate & Linear annealing 5e-2$\to$1e-2\\
    \hline
    \emph{MPC-Tracking (w/ reg)}\\
    Regularization coefficient $\zeta$ & 50\\
    \hline
    \end{tabular}
    \end{center}
    \vspace{-0.1in}
\end{table}

\section*{Appendix}
\subsection{Proof of Theorem 1}





\textit{Regularity conditions:} $p(s'|s,a)$, $\pi_\theta(s)$, $r(s,a)$, $d_0(s)$ are continuous in all parameters and variables $s$, $a$, $s'$.

\begin{proof}
We first unroll $\nabla_\theta V^\beta(s)$ with the definition of value function and the zeroth-order gradient trick (\ref{eq:esgrad}):

\begin{align*}
&\nabla_\theta V^{\beta}(s)\\
&=\nabla_\theta\mathbb{E}_{\epsilon_1\sim\mathcal{N}(0,I)}[Q^\beta(s,\pi_{\theta+\sigma\epsilon_1}(s))]\\
&=\nabla_\theta\mathbb{E}_{\epsilon_1\sim\mathcal{N}(0,I)}\Big[r(s,\pi_{\theta+\sigma\epsilon_1}(s))\\
&\quad\quad\quad+\int_\mathcal{S}\gamma p(s'|s,\pi_{\theta+\sigma\epsilon_1}(s))V^\beta(s')\mathrm{d}s'\Big]\\
&=\frac{1}{\sigma}\mathbb{E}_{\epsilon_1\sim\mathcal{N}(0,I)}\Big[\big(r(s,\pi_{\theta+\sigma\epsilon_1}(s))\\
&\quad\quad\quad+\int_\mathcal{S}\gamma p(s'|s,\pi_{\theta+\sigma\epsilon_1}(s))V^\beta(s')\mathrm{d}s'\big)\epsilon_1\Big]\\
&\quad+\mathbb{E}_{\epsilon_1\sim\mathcal{N}(0,I)}\int_\mathcal{S}\gamma p(s'|s,\pi_{\theta+\sigma\epsilon_1}(s))\nabla_\theta V^\beta(s')\mathrm{d}s'\\
&=\frac{1}{\sigma}\mathbb{E}_{\epsilon_1\sim\mathcal{N}(0,I)}\left[Q^\beta(s,\pi_{\theta+\sigma\epsilon_1}(s))\epsilon_1\right]\\
&\quad+\mathbb{E}_{\epsilon_1\sim\mathcal{N}(0,I)}\int_\mathcal{S}\gamma p(s\rightarrow s', 1, \pi_{\theta+\sigma\epsilon_1})\nabla_\theta V^\beta(s')\mathrm{d}s'
\end{align*}




We continue to unroll the recursive representation of $\nabla_\theta V^\beta(s)$ step by step till infinity. Note that the behavior policy $\beta$ is exactly the stochastic version of $\pi_\theta$ with parameter noise $\epsilon$ sampled i.i.d. at every timestep, we can rewrite the formula in the following form:

\begin{align*}
&\nabla_\theta V^{\beta}(s)=\frac{1}{\sigma}\mathbb{E}_{\epsilon\sim\mathcal{N}(0,I)}\left[Q^\beta(s,\pi_{\theta+\sigma\epsilon}(s))\epsilon\right]\\
&\quad+\int_\mathcal{S}\gamma p(s\rightarrow s', 1, \beta)\frac{1}{\sigma}\mathbb{E}_{\epsilon\sim\mathcal{N}(0,I)}\left[Q^\beta(s',\pi_{\theta+\sigma\epsilon}(s'))\epsilon\right]\mathrm{d}s'\\
&\quad+\int_\mathcal{S}\gamma^2 p(s \rightarrow s', 2, \beta)\nabla_\theta V^\beta(s')\mathrm{d}s'=\cdots\\
&=\int_\mathcal{S}\sum_{t=0}^{\infty}\gamma^t p(s\rightarrow s', t, \beta)\frac{1}{\sigma}\mathbb{E}_{\epsilon\sim\mathcal{N}(0,I)}\left[Q^\beta(s',\pi_{\theta+\sigma\epsilon}(s'))\epsilon\right]\mathrm{d}s'
\end{align*}

Now we can derive the zeroth-order policy gradient by taking the expectation over initial state distribution:

\begin{align*}
\nabla_\theta J(\theta)
&=\nabla_\theta\mathbb{E}_{s\sim d_0}[V^{\beta}(s)]
=\int_\mathcal{S}d_0(s)\nabla_\theta V^{\beta}(s)\mathrm{d}s\\
&=\int_\mathcal{S}\int_\mathcal{S}\sum_{t=0}^{\infty}\gamma^t d_0(s)p(s\rightarrow s', t, \beta)\\
&\quad\quad\quad\frac{1}{\sigma}\mathbb{E}_{\epsilon\sim\mathcal{N}(0,I)}\left[Q^\beta(s',\pi_{\theta+\sigma\epsilon}(s'))\epsilon\right]\mathrm{d}s'\mathrm{d}s\\
&=\frac{1}{\sigma}\mathbb{E}_{s\sim d_{\beta}}\mathbb{E}_{\epsilon\sim\mathcal{N}(0,I)}[Q^\beta(s,\pi_{\theta+\sigma\epsilon}(s))\epsilon]
\end{align*}

\end{proof}





\bibliographystyle{ieeetr}
\bibliography{ref}

\begin{thebibliography}{10}

\bibitem{neumann2019data}
M.~Neumann-Brosig, A.~Marco, D.~Schwarzmann, and S.~Trimpe, ``Data-efficient
  autotuning with bayesian optimization: An industrial control study,'' {\em
  IEEE Transactions on Control Systems Technology}, vol.~28, no.~3,
  pp.~730--740, 2019.

\bibitem{wang2021learning}
Y.~Wang, S.~Jiang, W.~Lin, Y.~Cao, L.~Lin, J.~Hu, J.~Miao, and Q.~Luo, ``A
  learning-based automatic parameters tuning framework for autonomous vehicle
  control in large scale system deployment,'' in {\em 2021 American Control
  Conference (ACC)}, pp.~2919--2926, IEEE, 2021.

\bibitem{marco2016automatic}
A.~Marco, P.~Hennig, J.~Bohg, S.~Schaal, and S.~Trimpe, ``Automatic lqr tuning
  based on gaussian process global optimization,'' in {\em 2016 IEEE
  international conference on robotics and automation (ICRA)}, pp.~270--277,
  IEEE, 2016.

\bibitem{o2009handbook}
A.~O'dwyer, {\em Handbook of PI and PID controller tuning rules}.
\newblock World Scientific, 2009.

\bibitem{garriga2010model}
J.~L. Garriga and M.~Soroush, ``Model predictive control tuning methods: A
  review,'' {\em Industrial \& Engineering Chemistry Research}, vol.~49, no.~8,
  pp.~3505--3515, 2010.

\bibitem{xu2019automated}
J.~Xu, Q.~Luo, K.~Xu, X.~Xiao, S.~Yu, J.~Hu, J.~Miao, and J.~Wang, ``An
  automated learning-based procedure for large-scale vehicle dynamics modeling
  on baidu apollo platform,'' in {\em 2019 IEEE/RSJ International Conference on
  Intelligent Robots and Systems (IROS)}, pp.~5049--5056, IEEE, 2019.

\bibitem{lambert2020objective}
N.~Lambert, B.~Amos, O.~Yadan, and R.~Calandra, ``Objective mismatch in
  model-based reinforcement learning,'' in {\em Learning for Dynamics and
  Control}, pp.~761--770, PMLR, 2020.

\bibitem{bansal2017goal}
S.~Bansal, R.~Calandra, T.~Xiao, S.~Levine, and C.~J. Tomlin, ``Goal-driven
  dynamics learning via bayesian optimization,'' in {\em 2017 IEEE 56th Annual
  Conference on Decision and Control (CDC)}, pp.~5168--5173, IEEE, 2017.

\bibitem{nair2020goal}
S.~Nair, S.~Savarese, and C.~Finn, ``Goal-aware prediction: Learning to model
  what matters,'' in {\em International Conference on Machine Learning},
  pp.~7207--7219, PMLR, 2020.

\bibitem{gevers2005identification}
M.~Gevers, ``Identification for control: From the early achievements to the
  revival of experiment design,'' {\em European journal of control}, vol.~11,
  no.~4-5, pp.~335--352, 2005.

\bibitem{piga2019performance}
D.~Piga, M.~Forgione, S.~Formentin, and A.~Bemporad, ``Performance-oriented
  model learning for data-driven mpc design,'' {\em IEEE control systems
  letters}, vol.~3, no.~3, pp.~577--582, 2019.

\bibitem{freeman2019learning}
D.~Freeman, D.~Ha, and L.~Metz, ``Learning to predict without looking ahead:
  World models without forward prediction,'' {\em Advances in Neural
  Information Processing Systems}, vol.~32, 2019.

\bibitem{risi2019deep}
S.~Risi and K.~O. Stanley, ``Deep neuroevolution of recurrent and discrete
  world models,'' in {\em Proceedings of the Genetic and Evolutionary
  Computation Conference}, pp.~456--462, 2019.

\bibitem{lucchini2020torque}
A.~Lucchini, S.~Formentin, M.~Corno, D.~Piga, and S.~M. Savaresi, ``Torque
  vectoring for high-performance electric vehicles: an efficient mpc
  calibration,'' {\em IEEE Control Systems Letters}, vol.~4, no.~3,
  pp.~725--730, 2020.

\bibitem{shahriari2015taking}
B.~Shahriari, K.~Swersky, Z.~Wang, R.~P. Adams, and N.~De~Freitas, ``Taking the
  human out of the loop: A review of bayesian optimization,'' {\em Proceedings
  of the IEEE}, vol.~104, no.~1, pp.~148--175, 2015.

\bibitem{roveda2020two}
L.~Roveda, M.~Forgione, and D.~Piga, ``Two-stage robot controller auto-tuning
  methodology for trajectory tracking applications,'' {\em IFAC-PapersOnLine},
  vol.~53, no.~2, pp.~8724--8731, 2020.

\bibitem{salimans2017evolution}
T.~Salimans, J.~Ho, X.~Chen, S.~Sidor, and I.~Sutskever, ``Evolution strategies
  as a scalable alternative to reinforcement learning,'' {\em arXiv preprint
  arXiv:1703.03864}, 2017.

\bibitem{jain2021optimal}
A.~Jain, L.~Chan, D.~S. Brown, and A.~D. Dragan, ``Optimal cost design for
  model predictive control,'' in {\em Learning for Dynamics and Control},
  pp.~1205--1217, PMLR, 2021.

\bibitem{mania2018simple}
H.~Mania, A.~Guy, and B.~Recht, ``Simple random search of static linear
  policies is competitive for reinforcement learning,'' {\em Advances in Neural
  Information Processing Systems}, vol.~31, 2018.

\bibitem{jain2021pixels}
D.~Jain, K.~Caluwaerts, and A.~Iscen, ``From pixels to legs: Hierarchical
  learning of quadruped locomotion,'' in {\em Conference on Robot Learning},
  pp.~91--102, PMLR, 2021.

\bibitem{huang2021accelerated}
R.~Huang, Y.~Chen, T.~Yin, X.~Li, A.~Li, J.~Tan, W.~Yu, Y.~Liu, and Q.~Huang,
  ``Accelerated derivative-free deep reinforcement learning for large-scale
  grid emergency voltage control,'' {\em IEEE Transactions on Power Systems},
  vol.~37, no.~1, pp.~14--25, 2021.

\bibitem{sutton2018reinforcement}
R.~S. Sutton and A.~G. Barto, {\em Reinforcement learning: An introduction}.
\newblock MIT press, 2018.

\bibitem{matni2019self}
N.~Matni, A.~Proutiere, A.~Rantzer, and S.~Tu, ``From self-tuning regulators to
  reinforcement learning and back again,'' in {\em 2019 IEEE 58th Conference on
  Decision and Control (CDC)}, pp.~3724--3740, IEEE, 2019.

\bibitem{amos2018differentiable}
B.~Amos, I.~Jimenez, J.~Sacks, B.~Boots, and J.~Z. Kolter, ``Differentiable mpc
  for end-to-end planning and control,'' {\em Advances in neural information
  processing systems}, vol.~31, 2018.

\bibitem{agrawal2020learning}
A.~Agrawal, S.~Barratt, S.~Boyd, and B.~Stellato, ``Learning convex
  optimization control policies,'' in {\em Learning for Dynamics and Control},
  pp.~361--373, PMLR, 2020.

\bibitem{buskens2001sensitivity}
C.~B{\"u}skens and H.~Maurer, ``Sensitivity analysis and real-time optimization
  of parametric nonlinear programming problems,'' in {\em Online Optimization
  of Large Scale Systems}, pp.~3--16, Springer, 2001.

\bibitem{gros2019data}
S.~Gros and M.~Zanon, ``Data-driven economic nmpc using reinforcement
  learning,'' {\em IEEE Transactions on Automatic Control}, vol.~65, no.~2,
  pp.~636--648, 2019.

\bibitem{gros2021reinforcement}
S.~Gros and M.~Zanon.
\newblock Reinforcement learning based on mpc and the stochastic policy
  gradient method.
\newblock In \emph{2021 American Control Conference (ACC)}, pages 1947--1952.
  IEEE, 2021.

\bibitem{raffin2022smooth}
A.~Raffin, J.~Kober, and F.~Stulp, ``Smooth exploration for robotic
  reinforcement learning,'' in {\em Conference on Robot Learning},
  pp.~1634--1644, PMLR, 2022.

\bibitem{silver2014deterministic}
D.~Silver, G.~Lever, N.~Heess, T.~Degris, D.~Wierstra, and M.~Riedmiller,
  ``Deterministic policy gradient algorithms,'' in {\em International
  conference on machine learning}, pp.~387--395, PMLR, 2014.

\bibitem{schulman2015high}
J.~Schulman, P.~Moritz, S.~Levine, M.~Jordan, and P.~Abbeel, ``High-dimensional
  continuous control using generalized advantage estimation,'' {\em arXiv
  preprint arXiv:1506.02438}, 2015.

\bibitem{konda1999actor}
V.~R. Konda and V.~S. Borkar, ``Actor-critic--type learning algorithms for
  markov decision processes,'' {\em SIAM Journal on control and Optimization},
  vol.~38, no.~1, pp.~94--123, 1999.

\bibitem{ge2021numerically}
Q.~Ge, Q.~Sun, S.~E. Li, S.~Zheng, W.~Wu, and X.~Chen, ``Numerically stable
  dynamic bicycle model for discrete-time control,'' in {\em 2021 IEEE
  Intelligent Vehicles Symposium Workshops (IV Workshops)}, pp.~128--134, IEEE,
  2021.

\bibitem{li2010model}
S.~Li, K.~Li, R.~Rajamani, and J.~Wang, ``Model predictive multi-objective
  vehicular adaptive cruise control,'' {\em IEEE Transactions on control
  systems technology}, vol.~19, no.~3, pp.~556--566, 2010.

\bibitem{nevergrad}
J.~Rapin and O.~Teytaud, ``{Nevergrad - A gradient-free optimization
  platform}.'' \url{https://GitHub.com/FacebookResearch/Nevergrad}, 2018.

\end{thebibliography}

\end{document}